\input{aipcheck}

\documentclass[
    ,final            
  ]
  {aipproc}

\layoutstyle{6x9}

\def\farcs{\hbox{$\,.\!\!^{''}$}}
\def\fdg{\hbox{$\,.\!\!^{\circ}$}}

\begin{document}

\title{The broadband afterglow of GRB 030328}


\author{E. Maiorano}{
  address={INAF-IASF, Bologna, Italy}
}
\author{N. Masetti}{
  address={INAF-IASF, Bologna, Italy}
}
\author{E. Palazzi}{
  address={INAF-IASF, Bologna, Italy}
}
\author{S. Savaglio}{
  address={The Johns Hopkins Univ., Baltimore, USA}
}
\author{E. Rol}{
  address={Univ. of Leicester, UK}
}
\author{P.M. Vreeswijk}{
  address={ESO-Santiago, Chile}
}
\author{E. Pian}{
  address={INAF-Oss. Astron. Trieste, Italy}
}
\author{P.A. Price}{
  address={Univ. of Hawaii, Honolulu, USA}
}
\author{B.A. Peterson}{
  address={Australian National Univ., Weston, Australia}
}
\author{M. Jel\'{i}nek}{
  address={IAA-CSIC, Granada, Spain}
}
\author{S.B. Pandey}{
  address={IAA-CSIC, Granada, Spain}
}
\author{M.I. Andersen}{
  address={AIP, Potsdam, Germany}
}
\author{A.A. Henden}{
  address={US Naval Observatory, Flagstaff, USA}
}

\begin{abstract}
We here report on the photometric, spectroscopic and polarimetric monitoring 
of the optical afterglow of the Gamma-Ray Burst (GRB) 030328 detected by 
{\it HETE-2}. We found that a smoothly broken power-law decay provides the 
best fit of the optical light curves, with indices $\alpha_1$ 
= 0.76$\pm$0.03, $\alpha_2$ = 1.50$\pm$0.07, and a break at 
$t_b$ = 0.48$\pm$0.03 d after the GRB. 
Polarization is detected in the optical $V$-band, with $P$ = (2.4$\pm$0.6) 
\% and $\theta$ = 170$^\circ$$\pm$7$^\circ$. 
Optical spectroscopy shows the presence of two absorption systems at $z$ = 
1.5216 $\pm$ 0.0006 and at $z$ = 1.295 $\pm$ 0.001, the former likely 
associated with the GRB host galaxy. The X--ray-to-optical spectral 
flux distribution obtained 0.78 days after the GRB was best fitted using 
a broken power-law, with spectral slopes $\beta_{\rm opt}$ = 0.47$\pm$0.15 
and $\beta_{\rm X}$ = 1.0$\pm$0.2. 
The discussion of these results in the context of the `fireball 
model' shows that the preferred scenario is a fixed opening angle 
collimated expansion in a homogeneous medium.
\end{abstract}

\maketitle


\section{Introduction}

GRB 030328 was a long, bright GRB detected on 2003 Mar 28.4729 UT, by the 
FREGATE, WXM, and SXC instruments onboard {\it HETE-2}, and rapidly 
localized with sub-arcminute accuracy (Villasenor et al. 2003).
About $\sim$1 hour after the GRB, its optical afterglow has been detected 
by the 40-inch Siding Spring Observatory (SSO) telescope (Peterson \& 
Price 2003). A study of the X--ray afterglow of GRB 030328 was performed 
by Butler et al. (2005) using {\it Chandra} data.

We report here on the study of the optical afterglow emission of GRB 
030328 made, within the GRACE\footnote{GRB Afterglow Collaboration at ESO; 
see {\tt http://www.gammaraybursts.org/grace/}} collaboration, performed
with 7 different optical telescopes.
A more detailed presentation of these data will appear in Maiorano et 
al. (2005).

\section{Observations}

Optical $UBVRI$ data of the GRB 030328 Optical Transient (OT), for a total 
of 130 photometry points, were acquired at the 40-inch SSO (Australia), 1m 
ARIES (India), 2.5m NOT (Spain), 1.54m ESO Danish, 2.2m ESO/MPG, ESO 
VLT-$Antu$ (Chile) and 1m USNO-FS (USA) telescopes. 

A series of six 10-min optical spectra was obtained at ESO-Paranal with 
VLT-{\it Antu} starting 0.59 d after the GRB. The Grism 300V was used with 
a nominal spectral coverage of 3600--8000 \AA~and a spectral dispersion of 
2$\farcs$7 \AA/pixel.

Linear polarimetry $V$-band observations were acquired between 0.66 and 
0.88 d after the GRB at VLT-{\it Antu}. Five complete imaging polarimetry 
cycles were performed.


\begin{figure}
\hspace{.2cm}
  \includegraphics[height=.25\textheight]{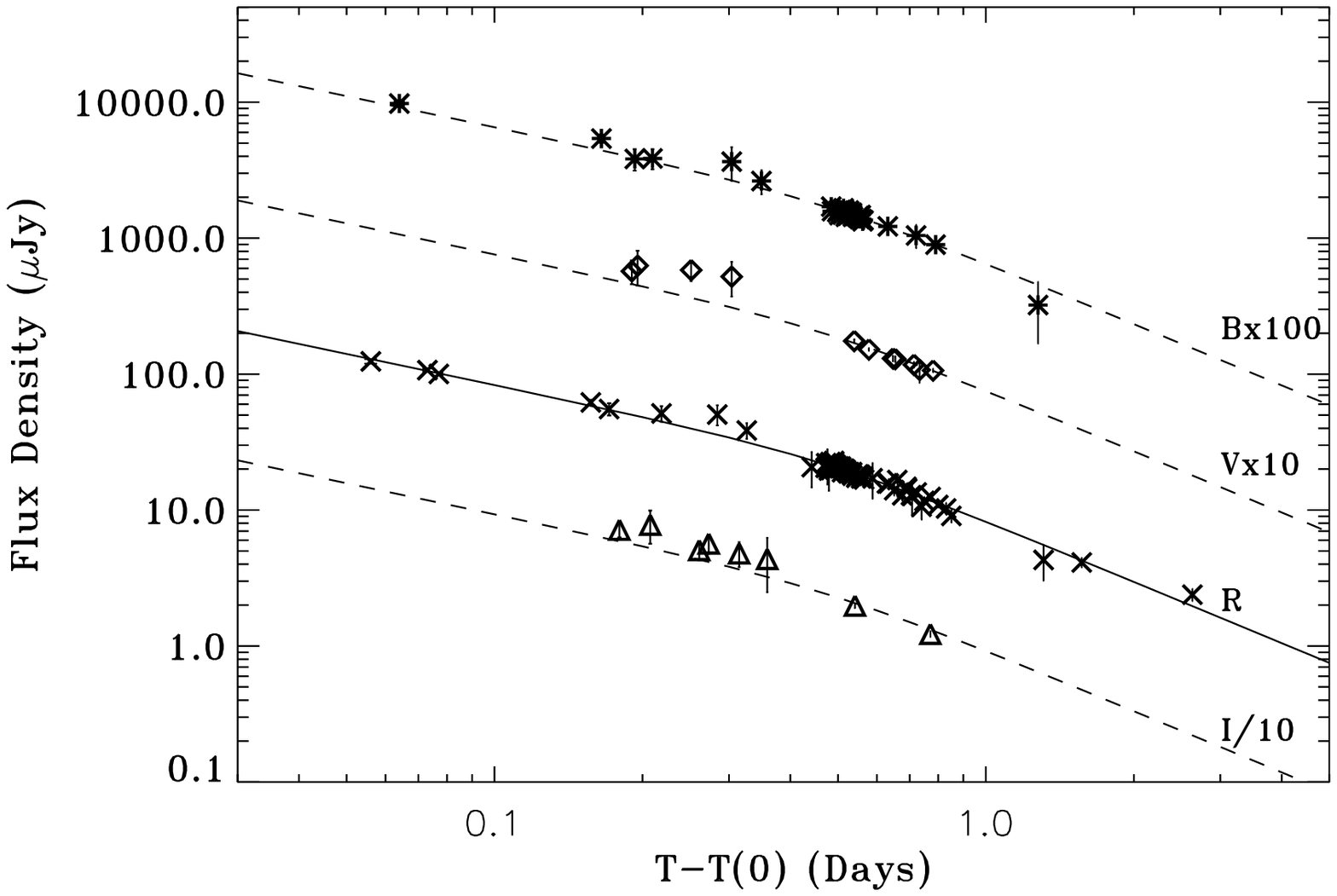}
\hspace{-.2cm}
  \includegraphics[height=.25\textheight]{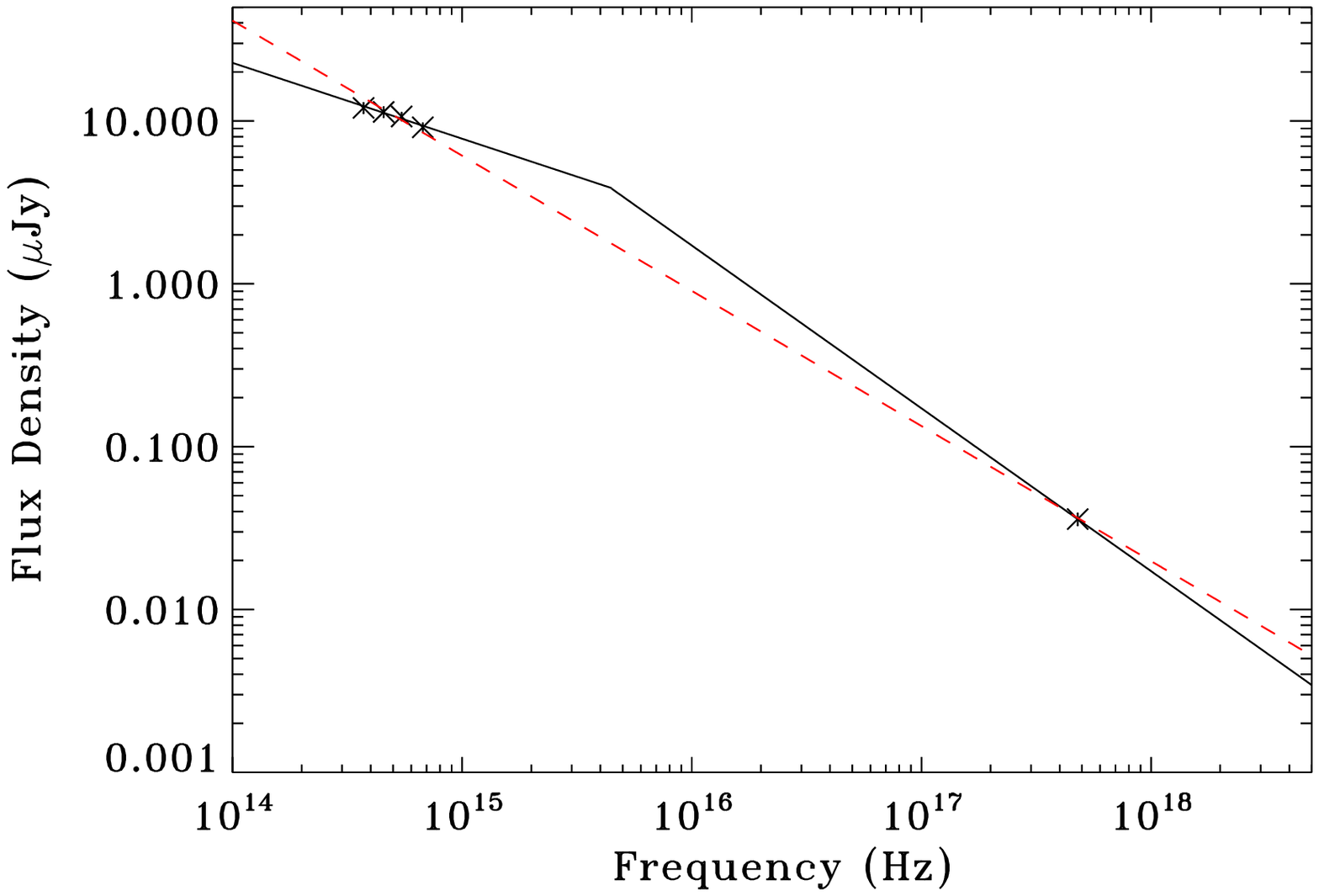}
\vspace{-2.5cm}
  \caption{{\it Left panel}: $BVRI$ light curves of the GRB 030328 
afterglow. The solid and dashed lines represent the best-fit of
the optical light curves. {\it Right panel}: broadband afterglow spectrum 
at 0.78 d after GRB 030328. The dashed line corresponds to a single 
optical-X--rays powerlaw spectral slope; the broken solid line indicates 
the description with $\nu_{\rm c}$ between the optical and the X--ray 
ranges.}
\end{figure}

\section{Results}

\subsection{Photometry}

In Fig. 1 (left) we plot our photometric measurements together with 
those reported in the GCN circulars\footnote{{\tt 
http://gcn.gsfc.nasa.gov/gcn/gcn3\_archive.html}}. For the cases in which 
no error was reported, a 0.3 mag uncertainty was assumed. The $UBVRI$ 
zero-point calibration was performed using the photometry by Henden (2003).

The optical data of Fig. 1 (left) were corrected for the Galactic 
foreground reddening assuming $E(B-V)$ = 0.047 mag (Schlegel et al. 1998). 
The GRB 030328 host galaxy emission in the $BVRI$ bands was computed from 
the data of Gorosabel et al. (2005) and subtracted from our optical data 
set.

The best fit of the $R$-band data (in Fig. 1, left) 
is obtained using a smoothly broken powerlaw (Beuermann et al. 1999), with 
temporal indices $\alpha_1 = 0.76 \pm 0.03$ and $\alpha_2 = 1.50 \pm 0.07$ 
before and after a break occurring at $t_b = 0.48 \pm 0.03$ d from the GRB 
trigger, and with $s = 4.0 \pm 1.5$ the parameter modeling the slope 
change rapidity. This best-fit curve describes well the data in the other 
bands also (see Fig. 1, left). This means that the decay of the OT can be 
considered as achromatic.

From the measured jet break time we can compute the jet opening angle
value for GRB 030328 which is, following Sari et al. (1999), $\theta_{jet}
\sim 3\fdg2$.

\subsection{Broadband analysis}

By using the available information, we have constructed the 
optical-to-X--ray spectral flux distribution of the GRB 030328 
afterglow at the epoch 0.78 days after the GRB, that is, the time with 
best broadband photometric coverage.

As Fig. 1 (right) shows, the best descriptions are a single powerlaw 
(dashed line) with a spectral index $\beta_{\rm X-opt}$ = 0.83$\pm$0.01, 
or a broken powerlaw with $\beta_{\rm opt}$ = 0.47$\pm$0.15 and assuming 
$\beta_{\rm X}$ = 1.0$\pm$0.2 from Butler et al. (2005).

However, the single powerlaw description of the broadband afterglow does 
not fit any of the synchrotron fireball scenarios (Sari et al. 1998, 
1999). Instead, in the broken powerlaw description, which means that the 
synchrotron cooling frequency $\nu_{\rm c}$ lies between the optical and 
X--ray bands, we obtain that the GRB 030328 afterglow broadband evolution 
is consistent with a jet-collimated expansion in a homogeneous medium with 
fixed opening angle (M\'esz\'aros \& Rees 1999) and with an electron 
distribution index $p$ = 2.

Assuming a negligible host absorption and using the optical and X--ray 
spectral slopes above, we obtain for $\nu_{\rm c}$ the value $5.9 \times 
10^{15}$ Hz, which places this frequency in the ultraviolet band.

\subsection{Spectroscopy}

Figure 2 shows the spectrum of the GRB 030328 OT. Most of the
significant features can be identified with Fe {\sc ii}, Mg {\sc ii},
Al {\sc ii} and C {\sc iv} absorption lines in a
system at a redshift of $z$ = 1.5216 $\pm$ 0.0006. These lines are
associated with the circumburst gas or interstellar medium in the
GRB host galaxy. A lower redshift absorption system at $z$ = 1.295
$\pm$ 0.001 is also found: for it, only two lines can be identified
(Fe {\sc ii} $\lambda$2600 and the unresolved Mg {\sc ii}
$\lambda$$\lambda$2796, 2803 doublet). Its detection indicates the
presence of a foreground absorber.

\subsection{Polarimetry}

After correcting for spurious field polarization, we found $Q_{\rm OT} = 
0.029 \pm 0.008$ and $U_{\rm OT} = -0.004 \pm 0.008$. The fit of the data 
with the relation of Di Serego Alighieri (1997) yielded for the OT a 
linear polarization $P = (2.4 \pm 0.6)$ \% and a polarization angle 
$\theta = 170^{\circ} \pm 7^{\circ}$, corrected for the polarization bias 
(Wardle \& Kronberg 1974).

In order to check whether variations of $P$ and $\theta$ occurred during 
the polarimetric run, we also separately considered each of the 5 single 
polarimetry cycles. Although with lower S/N, $P$ and $\theta$ are 
consistent with being constant across the whole polarimetric observation 
run.

\begin{figure}
  \includegraphics[height=.6\textheight, angle=270]{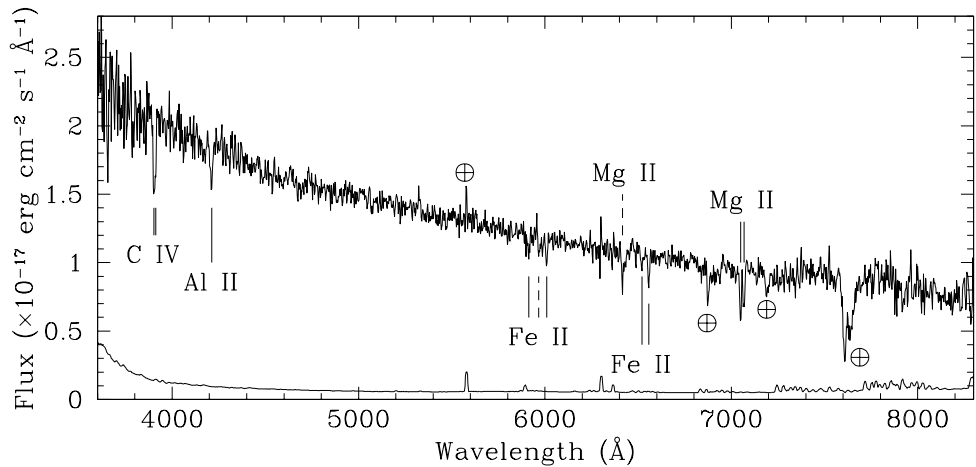}
  \caption{Optical spectrum (and corresponding Poisson error) of the 
afterglow of GRB 030328. Two systems at redshift $z$ = 1.5216 and $z$ 
= 1.295 are apparent: the former (solid hyphens) likely originates in the 
host galaxy, whereas the latter (dashed hyphens) is produced by a 
foreground absorber. The symbol $\oplus$ indicates atmospheric and 
telluric features.}
\end{figure}




\end{document}